\documentclass[aps,twocolumn,groupedaddress,showpacs]{revtex4}
\usepackage{amssymb, amsmath}
\usepackage{color}
\usepackage{graphicx}
\begin{document}
\title{Brownian motors in micro-scale domain: Enhancement of efficiency by noise}
%
%
\author{J. Spiechowicz$^1$, P. H\"anggi$^{2,3}$, and J. \L uczka$^{1,4}$\\
$^1$Institute of Physics, University of Silesia, 40-007 Katowice, Poland\\
$^2$Institute of Physics, University of Augsburg, 86135 Augsburg, Germany\\
$^3$Nanosystems Initiative Munich, Schellingstr, 4, D-80799 M\"unchen, Germany\\
$^4$Silesian Center for Education and Interdisciplinary Research, University of Silesia, 41-500 Chorz{\'o}w, Poland} 

%
\begin{abstract}
We study  a noisy drive mechanism for efficiency enhancement of Brownian motors operating on the micro-scale domain. 
It was proven [J. Spiechowicz {\it et al.},  J. Stat. Mech. \textbf{P02044}, (2013)] that biased noise $\eta(t)$ can induce normal and anomalous transport  processes similar to those  generated by a static force $F$  acting on inertial Brownian particles in a \emph{reflection-symmetric} periodic structure in presence of {\it symmetric}   unbiased time-periodic driving. Here, we show that within selected parameter regimes, noise $\eta(t)$ of the mean value $\langle \eta(t) \rangle = F$ can be significantly more effective than the deterministic force $F$: the motor can move much faster, its velocity fluctuations are much smaller and the motor efficiency increases  several times. These features hold true in both normal and  absolute negative mobility regimes. We demonstrate this with detailed simulations by resource to generalized white Poissonian noise. Our theoretical results can be tested and corroborated experimentally by use of a setup that consists of a resistively and capacitively shunted Josephson junction.
 The suggested strategy to replace $F$ by $\eta(t)$ may provide a new operating principle in which micro- and nanomotors could be powered by biased noise. 

\end{abstract}
\maketitle

\section{Introduction} 

Transport occurring in the micro-scale domain is strongly influenced by fluctuations and random perturbations. In certain regimes they  can play a dominant role. The typical situation is then that randomness hampers directed transport with respect to quantifiers such as the average transport velocity with particle motions being erratic and thus uncontrollable. However, a constructive role of both equilibrium and non-equilibrium fluctuations has since been demonstrated for many situations with the occurrence of several intriguing, noise assisted phenomena such as Brownian ratchets \cite{PH1}, stochastic resonance \cite{gama}, molecular motors and machines \cite{schilwa,motors}, genetic and biochemical regulatory systems \cite{steuer}, intracellular transport \cite{paul}, energy transport \cite{torres}, to mention a few. Fluctuations and noise may enhance the average velocity, reverse the natural transport direction or induce anomalous transport processes.
The conventional way to transport particles into a desired direction is to apply a constant  force $F$ pointing in this direction. Here we consider transport in spatially periodic systems and study a class of systems where both normal and anomalous transport regimes exist. We show that a stochastic force $\eta(t)$ of equal mean value as the deterministic force $F$ proves to be more effective than the deterministic counterpart.
Our idea and main message is: replace the deterministic forces by
suitable noise which in some regimes can appear to be much more effective. The proposal is in some sense
universal and can be realized both in classical and quantum systems; in condensed matter
physics and soft matter physics; in physical and biological systems. Examples where this idea could  be realized are: motors which are cold atoms in optical lattices \cite{renzo},  carbon nanotube motors based upon the torque generated by a flux of electrons
passing through a chiral nanotube \cite{lamb},  motors based on the chaotic
quantum dots \cite{raul}. 

The paper is organized as follows. In Sec. II we describe a mathematical model of the inertial Brownian motor which is driven by a time-periodic force and a constant force $F$ or  biased noise $\eta(t)$. The model has been previously studied in  various aspects and is proved to exhibit a rich diversity of anomalous transport characteristics  \cite{machura2008,machura2007,speer2007,spiechowicz2013}.  Sec. III contains a detailed analysis of three quantifiers characterizing transport processes, namely a long-time  averege velocity, its fluctuations and efficiency.  Sec. IV provides summary and some conclusions. 


\section{Model of Brownian motor} 

Modeling systems and understanding their generic properties discloses which components of the setup are crucial and which elements may be sub-relevant. Here we demonstrate this with an archetype class of  Brownian motors which is composed of a minimal number of elements but nevertheless is able to exhibit a wide class of anomalous transport features like:  absolute negative mobility (ANM) in a linear response regime,  negative mobility in a nonlinear response regime  and negative differential mobility \cite{machura2008}. The modeling uses a classical Brownian particle moving in a one dimensional periodic potential landscape. Using dimensionless variables the model consists of the following parts \cite{machura2007,speer2007}: (i) a particle of mass $M=1$, (ii) moving in a \textit{symmetric}, spatially periodic potential $V(x) = V(x + 1) = \sin (2\pi x)$ of period $L=1$, (iii) being driven by an unbiased time-periodic force $a\cos{\omega t}$ with amplitude $a$ and angular frequency $\omega$, and (iv) subjected to a constant force $F$. We also consider the counterpart by replacement of the force $F$ with biased noise $\eta(t)$. In order to make a comparison with the case of the deterministic force $F$, we set the mean value of the random force $\eta(t)$ equal to $F$, namely $\langle \eta(t) \rangle = F$. The corresponding dimensionless Langevin equations therefore read:
\begin{eqnarray}
	\label{eq1}
	\ddot{x} + \gamma\dot{x} = -V'(x) + a\cos(\omega t) + \sqrt{2\gamma D_T}\,\xi(t) + F,\\
	\label{eq2}
	\ddot{x} + \gamma\dot{x} = -V'(x) + a\cos(\omega t) + \sqrt{2\gamma D_T}\,\xi(t)+ \eta(t)\;.
\end{eqnarray}
Here, the dot and the prime denote a differentiation with respect to time $t$ and the Brownian particle's space coordinate $x$, respectively. The parameter $\gamma$ characterizes the friction coefficient. Thermal noise due to the coupling of the particle with thermostat is modeled by symmetric and unbiased $\delta$-correlated Gaussian white noise $\xi(t)$ with $\langle \xi(t) \rangle = 0$ and $\langle \xi(t)\xi(s) \rangle = \delta(t-s)$. Its intensity $D_T \propto k_B T$ is proportional to the thermal energy, where $T$ is the bath temperature with $k_B$ the Boltzmann constant. The dimensional version of Eq. (1) and corresponding scalings of length and time, etc., is detailed in Ref. \cite{machura2007} for an interested reader. From the symmetry property it follows that $F\to -F$ implies $\dot x(t) \to -\dot x(t)$ and in the following we assume a positive valued force $F>0$.
\begin{figure}[!t]
	\centering
	\includegraphics[width=0.9\linewidth]{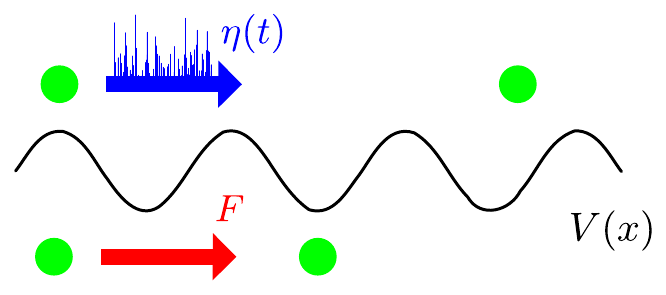}
\caption{Brownian motors moving in symmetric periodic structures in presence of an unbiased harmonic force $a\cos{\omega t}$ and driven by a static, biasing force $F$ can be transported much faster and in a more effective way when $F$ is replaced by noise $\eta(t)$ of equal average bias $\langle \eta(t) \rangle = F$.}
	\label{fig1}
\end{figure}
As a model of biased noise $\eta(t) > 0$ we choose a random sequence of $\delta$-pulses with random amplitudes defined in terms of \textit{generalized white Poissonian noise} \cite{hanggi19xx}
\begin{equation}
	\label{eq3}
	\eta(t) = \sum_{i=1}^{n(t)}z_i \delta(t - t_i)\;,
\end{equation}
where $t_i$ are the arrival times of Poissonian counting process $n(t)$ with Poisson parameter $\lambda$, i.e. the probability that $k$ pulses occur in the interval $(0,t)$ is given by the Poisson distribution $\mbox{Pr}\{n(t) = k \} = (\lambda t)^k \exp(-\lambda t)/k!$, with $\lambda$ being the mean number of $\delta$-pulses per unit time. The amplitudes $\{z_i\}$ of the pulses are mutually independent random variables of a common density $\rho(z)$ and are independent of the counting process $n(t)$. They are assumed to be exponentially distributed; i.e., $\rho(z) = \zeta^{-1} \theta(z) \exp(-z/\zeta)$, where $\theta(z)$ is the Heaviside step function. In consequence, all amplitudes $\{z_i\}$ are \textit{positive} and realizations of the process $\eta(t)$ are \textit{non-negative}, i.e. $\eta(t) \geq 0$. This biased non-equilibrium noise thus has a finite mean $\langle \eta(t) \rangle = \lambda\langle z_i \rangle = \sqrt{\lambda D_P }$ with co-variance, $(\langle \eta(t)\eta(s) \rangle - \langle \eta(t) \rangle \langle \eta(s) \rangle) = 2D_P \delta(t-s)$. We introduced the Poissonian noise intensity $D_P = \lambda \langle z_i^2 \rangle / 2 = \lambda \zeta^2$, where $\langle z_i^k \rangle = k!\zeta^k $ are the statistical moments of the amplitudes $\{z_i\}$. We also assume that the thermal equilibrium fluctuations $\xi(t)$ are uncorrelated with non-equilibrium noise $\eta(t)$; i.e., $\langle \xi(t)\eta(s) \rangle = \langle \xi(t) \rangle \langle \eta(s) \rangle = 0$. The influence of the Poissonian noise parameters $\lambda$ and $D_P$ on stochastic realizations of $\eta(t)$ is presented in \cite{spiechowicz2013}. Here, we only mention two extreme regimes. The first limiting case is when both $\lambda$ and $D_P$ are large, then the particle is frequently kicked by large $\delta$-pulses. On the contrary, when both $\lambda$ and $D_P$ are small, then the particle it is rarely kicked by $\delta$-pulses of small amplitudes.

\section{Efficiency of  motor}

The most important quantity characterizing a Brownian motor is its directed average velocity. In the asymptotic long time regime it is determined by the relation \cite{gama,jung1990,jung1993}
\begin{equation}
	\label{eq4}
	\langle v \rangle = \lim_{t\to\infty} \frac{\omega}{2\pi} \int_{t}^{t+2\pi/\omega} {\mathbb E}[v(s)] \; ds,
\end{equation}
where $\mathbb E[v(t)]$ denotes the average of the actual velocity $v(t) =\dot x(t)$ over the noise realizations and initial conditions. Although this average velocity is a main quantifier for the transport it is, however, not necessarily of decisive character in attaining an optimal efficiency for the working operation. For example, a large transport velocity is of little use if the fluctuations are too erratic around the average velocity, thus spoiling effectiveness. We next study the size of the velocity fluctuations. In the long time regime these are given by
\begin{equation}
	\label{eq5}
	\sigma_v^2 = \langle v^2 \rangle - \langle v \rangle^2.
\end{equation}
Typically the velocity mainly assumes values within the interval $v(t) \in \left(\langle v \rangle - \sigma_v, \langle v \rangle + \sigma_v\right)$. If these fluctuations are very large, i.e. if $\sigma_v > \langle v \rangle$, it implies that the Brownian motor can move for some time in the direction opposite to its average velocity $\langle v \rangle$. As a measure of its effectiveness we consider a common measure, namely its Stokes \textit{efficiency} $\varepsilon_S$ \cite{wang,wang2}: it is evaluated as the ratio of the dissipated power $F_v \langle v \rangle$, associated with the directional movement against the mean viscous force $F_v = \gamma \langle v \rangle$, to the input power $P_{in}$ \cite{machura20042010,PhysE}; i.e.,	
\begin{equation}
	\label{eq6}
	\varepsilon_S = \frac{F_v \langle v \rangle}{P_{in}} = \frac{\gamma \langle v \rangle^2}{P_{in}} = \frac{\langle v \rangle^2}{\langle v^2 \rangle - D_T}\;.
\end{equation}
Here, $P_{in}$ is supplied to the system by all external forces, i.e. by both, the ac-driving $a\cos(\omega t)$ and the static force $F$ or the random force $\eta(t)$. From an energy balance of the underlying equations of motion (\ref{eq1}) or (\ref{eq2}) it follows that $P_{in} = \gamma(\langle v \rangle^2 + \sigma_v^2 - D_T) = \gamma(\langle v^2 \rangle - D_T)$,which is always positive valued \cite{machura20042010,PhysE,jung11}. We note that this non-equilibrium efficiency (\ref{eq6}) does not coincide with the thermodynamic efficiency, e.g. see the discussion on p. 218 of Ref. \cite{schilwa}. Physical intuition tells us that a decrease of the variance $\sigma_v^2$ generates a smaller input power and hence to an increase of the overall efficiency. Put differently, transport is optimized in regimes which \textit{maximize} the directed velocity and \textit{minimize} its fluctuations.
\begin{figure*}[t]
	\centering
	\includegraphics[width=0.3\linewidth]{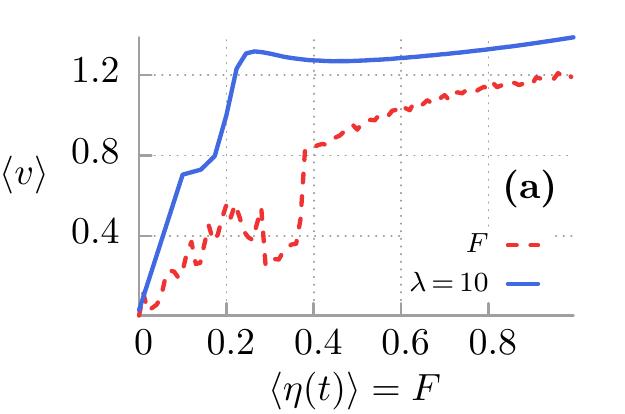}
	\includegraphics[width=0.3\linewidth]{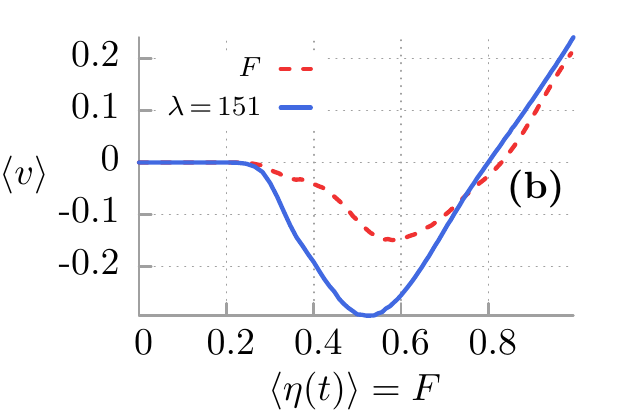}
	\includegraphics[width=0.3\linewidth]{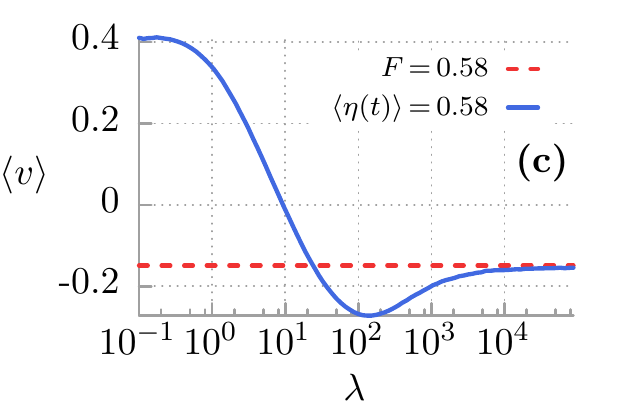}\\
	\includegraphics[width=0.3\linewidth]{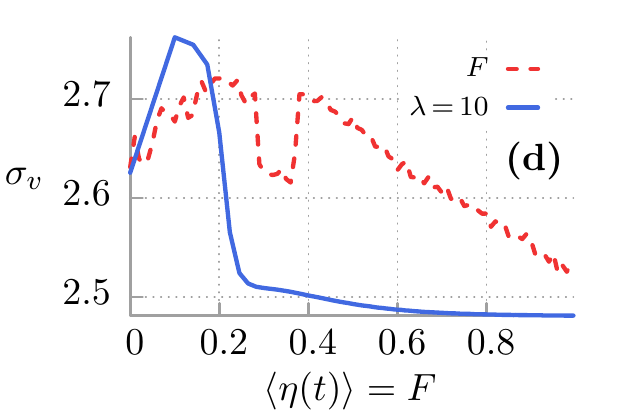}
	\includegraphics[width=0.3\linewidth]{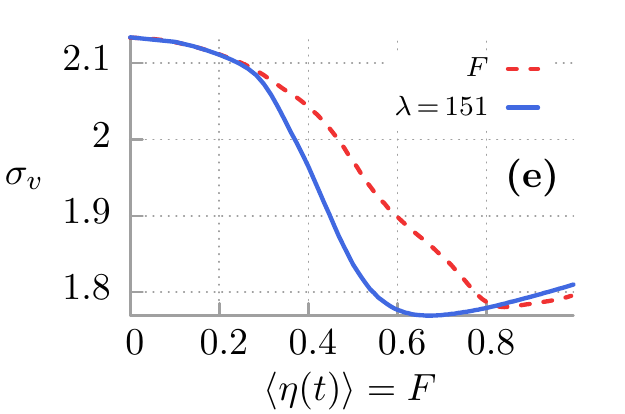}
	\includegraphics[width=0.3\linewidth]{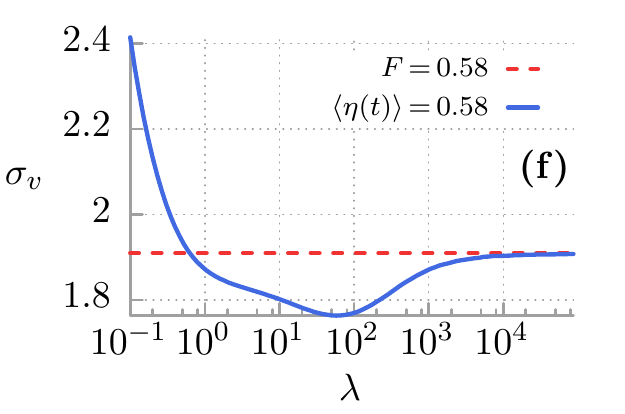}\\
	\includegraphics[width=0.3\linewidth]{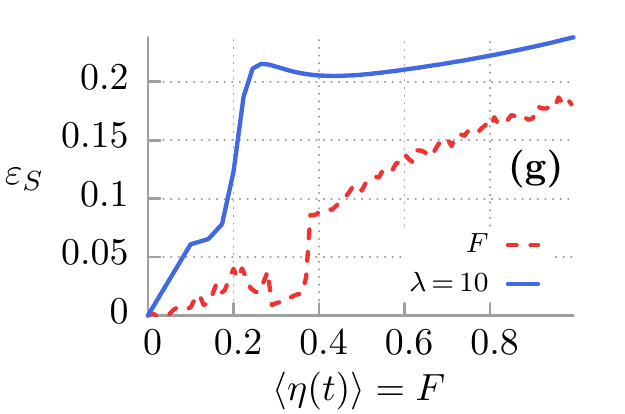}
	\includegraphics[width=0.3\linewidth]{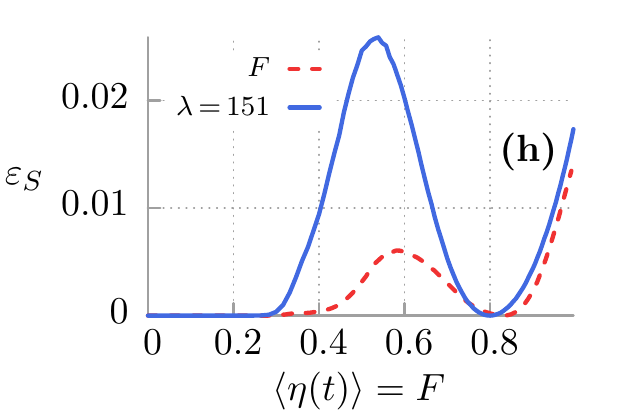}
	\includegraphics[width=0.3\linewidth]{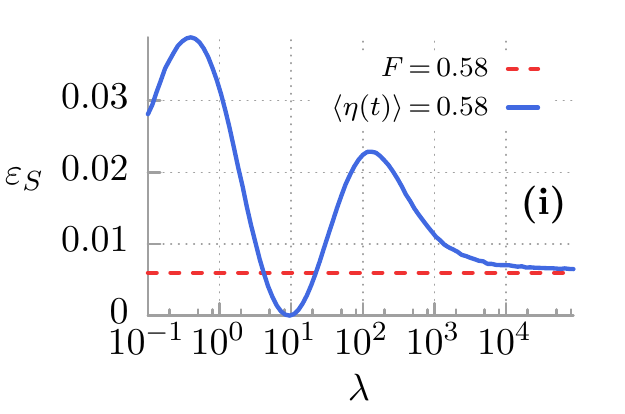}
\caption{Normal (left panels) and anomalous (middle panels) transport regimes. Upper horizontal panels: the asymptotic averaged velocity $\langle v \rangle$. Middle horizontal panels: the velocity fluctuations $\sigma_v$. Bottom horizontal panels: the Stokes efficiency $\varepsilon_S$. All as a function of the deterministic constant bias $F$ and the random force $\eta(t)$ modeled here by white Poissonian noise of the mean value $\langle \eta(t) \rangle = F$. In right panels,  the dependence  on the spiking rate $\lambda$ is shown for fixed mean value of  noise $\langle \eta \rangle = 0.58$.
Parameters in the left panels: $a = 27.5, \omega = 8.5, \gamma = 1.194, D_T = 10^{-6}, \lambda = 10$. In the middle and right panels they are as follows: $a = 8.95, \omega = 3.77, \gamma = 1.546, D_T = 0.001, \lambda = 151$.}
	\label{fig2}
\end{figure*}

The deterministic system dynamics, i.e. if $D_T = D_P = 0$ in (\ref{eq1}), is extremely rich in complexity \cite{jung1996mateos2000,mateos}. Particularly, main features of the asymptotic behavior are locked states in which the motion of the Brownian particle is bounded to one or several spatial periods, chaotic and running states in which movement is unbounded in space. The latter modes of transport are crucial for the occurrence of the deterministic directed transport. Adding the Poissonian noise $\eta(t)$ or thermal fluctuations $\xi(t)$ activates a stochastic dynamics for which transitions between neighboring states are induced and therefore can result in diffusive or even directed transport. Since the Fokker-Planck-Kolmogorov-Feller master equation \cite{hanggi19xx} corresponding to the white noise driven Langevin equations (\ref{eq1}) or (\ref{eq2}) surely cannot be solved analytically, we performed extensive numerical simulations. The specific details of the employed numerical code can be found in \cite{spiechowicz2013}. Here, we only mention that all numerical simulations were done by use of a CUDA environment which is implemented on a modern desktop GPU. This scheme allowed for a speed-up of a factor of the order $10^3$ as compared to a present-day CPU method \cite{januszewski2009}. Our so obtained main results are presented next.

\subsection{Normal and ANM regimes} 

 The vast majority of stable running states point into the positive direction for $F>0$. However, there are also running stable states which on average move in the opposite direction to $F>0$ and the phenomenon of ANM occurs \cite{machura2007}. It can been shown \cite{spiechowicz2013} that similar anomalous transport processes occur for the system (\ref{eq2}). 
 The  negative mobility is generated  by various
mechanisms.  In some regimes it is caused by
thermal equilibrium fluctuations \cite{machura2007} and ANM  is absent for
vanishing fluctuations. In other regimes, ANM can occur
in the deterministic system while noise either destroys the effect or diminishes its strength \cite{speer2007,machura2008}. Nevertheless, the origin of ANM is in the noise--free structure of stable and
unstable orbits, see a detailed discussion in Ref. \cite{Acta}. 

Not surprisingly, it is practically sheer impossible to probe numerically the full parameter space $\{\gamma, a, \omega, D_T, F, \lambda, D_P\}$. We have chosen our simulations in regimes which exhibit a most important and intriguing transport behavior. We start with the asymptotic average velocity. This transport velocity is reduced when the deterministic force $F$ acting on the Brownian motor is replaced by the biased noisy drive $\eta(t)$, as one naively would also expect. However, there are regimes in the parameters space where the use of a random force is more effective. In Fig. 2, we illustrate two specific cases. The average velocity is depicted as a function of the mean random force, i.e. the static bias, $\langle \eta(t) \rangle = F$. The left panel (a) corresponds to a normal transport regime with average velocity pointing into the direction of the bias. For this specific set of parameters, the average velocity as a function of the bias $F$ is jagged and not smooth. The reason is that the chosen temperature is very small and both, amplitude and frequency of the ac-driving are relatively high. On the other hand, the noisy force $\eta(t)$ causes a smoothing of $\langle v \rangle$ and in parallel it enhances its magnitude. In particular, for a bias $\langle \eta(t) \rangle \approx 0.3$ the random force increases the average velocity by a factor of four in comparison to the application of a deterministic force $F$.

In panel (b), the regime of absolute negative mobility (ANM) is depicted. The characteristic feature is the emergence of interval where for sufficiently large $\langle \eta(t) \rangle >0$ the random force can induce negative average velocity $\langle v \rangle <0$ . Moreover, there exists an optimal value for the bias $\langle \eta(t) \rangle \approx 0.58$ at which the average velocity takes its minimal value. Most interesting is the fact that in the case of the stochastic force the minimal value of $\langle v \rangle$ is nearly two times lower than in the corresponding deterministic case. The origin of this effect is illustrated in panel (c), where we depict the average velocity $\langle v \rangle$ as a function of the spiking rate $\lambda$ for fixed mean value $\langle \eta(t) \rangle = \sqrt{\lambda D_P } = 0.58$. We see that depending on the frequency $\lambda$ of $\delta$-pulses we are dealing with three qualitatively different regimes. For small $\lambda$ the Brownian motor is rarely kicked by large $\delta$-pulses (because the intensity $D_P$ is large). This situation corresponds to normal, Ohmic-like transport behavior. On the contrary, when $\lambda$ is sufficiently increased 
and the particle is very frequently kicked by small $\delta$-pulses. In this case the response of the system is anomalous and similar to the scenario when an equivalent static, deterministic force acts \cite{spiechowicz2013}. The most interesting effect takes place for moderate $\lambda$. Then the motor operates in the anomalous regime, moves in the negative direction and its average velocity is enhanced as compared to the case of the static force. There exists an optimal value of $\lambda$ for which this effect is most pronounced, namely $\lambda = 151$. Moreover, the spiking rate $\lambda$ also serves as a control parameter for the direction of motor movement.

\subsection{Velocity fluctuations} 

 Let us next discuss the relation between the average velocity and its fluctuations. This study is of relevance for the optimization of efficiency. In panels (d) and (e), the velocity variance $\sigma_v$ is depicted as a function of the mean value of the white Poissonian noise $\langle \eta(t) \rangle$ and the corresponding static force $F$. Upon closer inspection one observes that the velocity fluctuations undergo rapid changes, yielding a large variance for small bias. This statement holds true regardless of the choice of the symmetry breaking mechanism. Further increase of the random or deterministic bias leads to a decrease of the velocity fluctuations for both analyzed scenarios. However, when the motor is driven by the random force these fluctuations become reduced faster. This is clearly noticeable in the normal transport regime, see panel (d). The result seems to be counter-intuitive as one would expect that the random force should enhance fluctuations. Furthermore, as is shown in panel (e), there exists an optimal value for bias $\langle \eta(t) \rangle$ at which the velocity fluctuations assume a minimal value. One should notice that this minimum nearly coincides with value of $\langle \eta(t) \rangle$ for which the negative average velocity in the anomalous regime take their minimal values, cf. panel (b). This bring us to the conclusion that the replacement of the deterministic force with the white generalized Poissonian noise not only decreases the negative-valued average velocity but simultaneously also minimizes its fluctuation behavior.

Panel (f) illustrates the impact of the spiking frequency $\lambda$ on the velocity fluctuations for fixed mean $\langle \eta(t) \rangle = 0.58$ in the anomalous regime. When $\lambda$ is very small the Brownian particle is rarely kicked with large $\delta$-pulses and its velocity undergoes rapid changes yielding a large variance. The opposite situation takes place in the case of a very large spiking frequency $\lambda$. For small to moderate $\lambda$ the velocity fluctuations are smaller as compared to the case with the static force. Moreover, there exists an optimal value for the spiking frequency $\lambda$ at which the velocity fluctuations are minimal.

\subsection{Stokes efficiency} 

 Of main interest is the overall efficiency of the Brownian motor operation. In panels (g) and (h) we depict the Stokes efficiency $\varepsilon_S$ {\it vs.} the mean $\langle \eta(t) \rangle$ and $F$. From its definition, the Stokes efficiency approaches zero for small statistical or deterministic bias values $F$. In both, the normal and the anomalous transport regimes there occurs an optimal Stokes efficiency. By use of the biased random force, $\varepsilon_S$ this efficiency grows by a factor of $4$ over the value obtained with a deterministic force. This effect is directly related to the property that the stochastic force enhances the absolute value of velocity and also minimize its fluctuations. Further increase of the bias leads to a decrease of the efficiency of the occurring transport process. For very large $\langle \eta(t) \rangle$ it approaches the value 1.

Panel (i) depicts the dependence of $\varepsilon_S$ on the spiking rate $\lambda$ for fixed $\langle \eta(t) \rangle$. The reader can observe two peaks. The first one corresponds to the regime of small $\lambda$ which represents normal, Ohmic-like transport. This maximum is associated with the fact that in that case the Brownian motor moves with large positive average velocity and its fluctuations are moderate. The second peak is located in the ANM-transport regime. It occurs for medium spiking rates $\lambda$. When $\lambda$ is large, then the transport process induced by this non-equilibrium noise approaches the behavior of the deterministic drive. This result corroborates with the previous statement concerning the qualitative equivalence between the white Poissonian noise and the deterministic force for very large spiking rates.

\section{Summary} 

We have investigated two models of the inertial Brownian motors: one driven by the deterministic force $F$ and the other propelled by biased non-equilibrium noise $\eta(t)$. We find domains in the parameters space such that when $F$ is replaced by $\eta(t)$ of equal average bias, 
the motor velocity is several times greater, the velocity fluctuations are
reduced several times and its efficiency becomes several times enhanced within tailored parameter regimes, both in its normal and its absolute negative mobility regime. Specific results are detailed for generalized white Poissonian noise. The main conclusion, remain valid, however, as well for other models of random perturbations (not depicted). Thus, the idea that random biased forces can be beneficial over deterministic biasing carries potential for practical realization in physics of Brownian motors, For example, it can be validated by use of a setup consisting of a the resistively and capacitively shunted Josephson junction device operating in corresponding experimentally accessible regimes.
 An exemplary set of physical  parameters in the ANM regime can be similar as in the experiment \cite{nagel2008}. 
In order to evaluate this set, we follow the method described in Ref. {\cite{machura2008}.  For  operational temperature $T=4$K and the set of parameters presented in the middle panels of Fig.2,   the critical Josephson 
current $I_c \approx 170 \mu$A. For a realistic capacitance  $C=20$pF, 
the plasma frequency $\omega_p \approx 160$GHz. The amplitude of the ac  current  is   $I_a \approx 240 \mu$A, the ac-angular frequency  
is $\Omega \approx 96$GHz and the dc current or the mean value of the Poisson noise is $I_d \approx 15.5 \mu$A. Under these conditions, 
the negative-valued ANM voltage is $V \approx -15.7 \mu$V for the Poissonian noise case and $V \approx -7.9 \mu$V for the deterministic force. 

The proposed mechanism of a "reduction of noise by noise" may explain exotic transport phenomena not only in physical but also in biological settings and, additionally, can be implemented in enhancing the working efficiency of synthetic molecular motors, all of which {\it in situ} operate in strongly fluctuating environments.

\section*{Acknowledgments}
This work was supported by the MNiSW program ”Diamond Grant” (J. S.) and NCN grant DEC-2013/09/B/ST3/01659 (J. {\L}.)

\end{document}